\documentclass[proceedings,letterpaper,10pt]{IEEEtran}

\usepackage{lipsum}
\usepackage{tikz}
\usepackage{amsmath}

\usepackage{wrapfig}
\usepackage{longtable,tabu}
\usepackage{booktabs}
\usepackage{threeparttable} 
\usepackage{subcaption}
\usepackage{enumitem}

\usepackage{adjustbox}
\usepackage{array}
\usepackage{multirow}

\newcommand*{\setuptable}{
	\renewcommand{\arraystretch}{1.1}
	\setlength{\arrayrulewidth}{0.1em}
	\setlength\tabcolsep{3.75pt}
	\centering
}

\newcommand*{\headrow}[1]{\multicolumn{1}{c}{\adjustbox{angle=45,lap=\width-0.5em}{#1}}}
\newcommand*{\headline}[1][3cm]{\adjustbox{angle=45,lap=\width-.3mm}{\rule{#1}{0.075em}}}

\usepackage{wasysym}
\newcommand*{\full}{\CIRCLE}

\newcommand*{\none}{\Circle}

\usepackage{csquotes}

\makeatletter
\renewcommand\@makefntext[1]{\leftskip=2em\hskip-.5em\@makefnmark#1}
\makeatother

\usepackage{pifont}
\newcommand*{\good}{{\textcolor[RGB]{112,173,71}{\ding{51}}}}
\newcommand*{\bad}{{\textcolor[RGB]{192,0,0}{\ding{55}}}}

\begin{document}

%don't want date printed
\date{}

% Six Navy SOC Analysts Usage and Perceptions of Two ML Tools
%\title{\Large \bf Usability and Adoption of ML-Based Cybersecurity Tools  \\
  %A Qualitative Analysis at the National Cyber Range}
  
\title{\Large \bf  An Assessment of the Usability of Machine Learning \\ Based Tools for the Security Operations Center}

% if you leave this blank it will default to a possibly ugly attempt 
% to make the contents of the \author command below into a string
\def\plainauthor{Author name(s) for PDF metadata. Don't forget to anonymize for submission!}

\author{\IEEEauthorblockN{Sean Oesch\IEEEauthorrefmark{1},
Robert Bridges\IEEEauthorrefmark{1}, Jared Smith\IEEEauthorrefmark{1}, Justin Beaver\IEEEauthorrefmark{1}\\John Goodall\IEEEauthorrefmark{1}, Kelly Huffer\IEEEauthorrefmark{1}, Craig Miles\IEEEauthorrefmark{2}, Dan Scofield\IEEEauthorrefmark{2}\\}
\IEEEauthorblockA{\IEEEauthorrefmark{1}Oak Ridge National Laboratory, \IEEEauthorrefmark{2}Assured Information Security\\
oeschts@ornl.gov}
\IEEEcompsocitemizethanks{\IEEEcompsocthanksitem Notice: This manuscript has been authored by UT-Battelle, LLC under Contract No. DE-AC05-00OR22725 with the U.S. Department of Energy. The United States Government retains and the publisher, by accepting the article for publication, acknowledges that the United States Government retains a non-exclusive, paid-up, irrevocable, world-wide license to publish or reproduce the published form of this manuscript, or allow others to do so, for United States Government purposes. The Department of Energy will provide public access to these results of federally sponsored research in accordance with the DOE Public Access Plan (http://energy.gov/downloads/doe-public-access-plan).}
}
%for single author (just remove % characters)
%\author{
%{\rm Author names redacted}\\
%Author affiliations redacted
%\and
%{\rm Second Name}\\
%Second Institution
% copy the following lines to add more authors
% \and
% {\rm Name}\\
%Name Institution
%} % end author
% AIS Authors:
%  Craig Miles / Assured Information Security / craig@craigmil.es
%  Daniel Scofield / Assured Information Security / dan@dscofield.com
%  Patrick McHarris / Assured Information Security / mcharrisp@ainfosec.com

\maketitle

% \thecopyright %% creating an error so removed

%The paper abstracts should contain a sentence summarizing the contribution to the field and literature.
\begin{abstract}
Gartner, a large research and advisory company, anticipates that by 2024 80\% of security operation centers (SOCs) will use machine learning (ML) based solutions to enhance their operations.
In light of such widespread adoption, it is vital for the research community to identify and address usability concerns.
This work presents the results of the first in situ usability assessment of ML-based tools.
With the support of the US Navy, we leveraged the national cyber range---a large, air-gapped cyber testbed equipped with state-of-the-art network and user emulation capabilities---to study six US Naval SOC analysts' usage of two tools.
Our analysis identified several serious usability issues, including multiple violations of established usability heuristics for user interface design.  
We also discovered that analysts lacked a clear mental model of how these tools generate scores, resulting in mistrust and/or misuse of the tools themselves.
Surprisingly, we found no correlation between analysts' level of education or years of experience and their performance with either tool, suggesting that other factors such as prior background knowledge or personality play a significant role in ML-based tool usage.
Our findings demonstrate that ML-based security tool vendors must put a renewed focus on working with analysts, both experienced and inexperienced, to ensure that their systems are usable and useful in real-world security operations settings. 
\end{abstract}

\section{Introduction}

Security operation centers (SOCs)\textemdash teams of security analysts who continually guard networks against cyber attacks\textemdash now employ widespread data collection capabilities \cite{bridges2018information} and follow a ``defense in depth'' strategy~\cite{colarik2015establishing,tirenin1999concept} that includes a tapestry of tools for blocking, alerting, logging, and providing situational awareness. 
To effectively defend networks and allow analysts to gain actionable insights from this wealth of SOC data, a robust research community and a burgeoning cyber tech industry are integrating machine learning (ML) into novel solutions.
Common categories of tools integrating ML to effectively leverage SOC data include the following: modern endpoint protection/anti-virus (AV), endpoint detection and response (EDR), network situational awareness/anomaly detection (AD), user and entity behavioral analytics (UEBA), security incident and event management (SIEM) systems, and security orchestration and automated response (SOAR).

Gartner anticipates that by 2024 80\% of SOCs will use ML-based tools to enhance their operations.
In light of such widespread adoption, it is vital for the research community to both enumerate and address usability concerns.
While prior work has sought to understand the issues that plague SOC operations ~\cite{kokulu2019matched,bridges2018information,goodall2004work,botta2007towards} and create more effective ML tools for SOCS~\cite{arendt2015ocelot,goodall2018situ,best2010real,sopan2018building}, no prior work examines analysts' usage of ML-based tools in situ. 
This gap in the research is understandable because it is non-trivial to gain access to a high fidelity testing environment and recruit actual SOC analysts to participate in such a study.

In this work, we share the results of an in situ study made possible by our sponsor, the US Navy, who purchased time at a testing center known for conducting high fidelity cyber events\textemdash the National Cyber Range (NCR) in Orlando, Florida.
The Navy also provided six analysts from their SOCs to participate in the study.
With these resources at our disposal, we designed a test to identify potential usability issues in two ML-based tools\textemdash one AV tool that carved files out of network traffic and a real time network-level AD tool. 

We configured the NCR to simulate a network with $\sim$1000 IPs that included emulated users with access to email, social media, and general websites, as well as management infrastructure and an out-of-band network allowing analysts to access the technologies under evaluation. 
We then conducted red team campaigns against the network, one for each tool, and observed analysts as they interacted with the tools.
After testing, we asked analysts to complete a follow-up survey and discussed their experiences in a focus group.

Our analysis identified several serious usability issues, including multiple violations of established usability heuristics for user interface design.  
We also discovered that analysts lacked a clear mental model of how these tools generate scores, resulting in mistrust and/or misuse of the tools themselves.
Surprisingly, we found no correlation between analysts' level of education or years of experience and their performance with either tool, suggesting that other factors such as prior background knowledge or personality play a significant role in ML-based tool usage.
Our findings demonstrate that ML-based security tool vendors must put a renewed focus on working with analysts, both experienced and inexperienced, to ensure that their systems are usable and useful in real-world security operations settings.

\section{Background}{\label{sec:background}}
In this section, we describe the testbed where we conducted the evaluation and the two tools tested, as well as providing an overview of related work.

\subsection{National Cyber Range}

The National Cyber Range (NCR)~\cite{ferguson2014national} provided the high-fidelity environment for our study.
The NCR is a large, air-gapped cyber testbed equipped with state-of-the-art network and user emulation capabilities that enables the rapid emulation of complex, operationally representative networks that can scale to over 50,000 virtual nodes.  
The range included ``user machines'', emulating real users, a management network with services such as DNS and Active Directory, a server network with on-premise servers such as Apache and IIS, and an "external network" for email, social media, and general websites.
The technologies under test were all connected to a core router and/or to a passive tap so each had access to all network traffic and could communicate with any host-based clients forwarding data.
User terminals connected to the two technologies under test via an out-of-band network and allowed evaluation team members and/or security analysts (users) access to the user interface (UI). 
%The technologies under test were only accessible via this out-of-band network, so they were effectively transparent and inaccessible by the rest of the virtualized environment.

\subsection{Tools Tested}
\label{sec:tools} 

This study included two tools, a commercially available network-based malware detection tool and a government off-the-shelf, anomaly detection tool. Because of a non-disclosure agreement, we cannot disclose the name of the vendor who supplied the first tool. 
It is a network-based, static-analysis, malware detection tool (NSDT) that is capable of identifying both existing and new/polymorphic attacks in near real time using an on-premises (on-prem) appliance to passively monitor network traffic. 
The technology centers on a binary (benign/malicious) classification of files and code snippets extracted from network traffic.

The second tool, Situ, is a government off-the-shelf (GOTS) tool for near real time network-level anomaly detection and situational awareness/exploration through visualization \cite{goodall2018situ}. 
Overall, the tool identifies anomalous\textemdash not necessarily malicious\textemdash network behavior and provides an interface for situational awareness, hunting, and forensic investigation. 
The system ingests network flows, the metadata of IP-to-IP communication and/or firewall logs.

\subsection{Related Work}
\label{sec:related-works} 
Related works fall into four categories---visual analytics to aid security analysts, methods to evaluate the effectiveness of security tools in the context of a SOC, studies on SOC operations, and ML for cybersecurity. 
While prior work relied heavily on interviews or surveys for data collection, our work represents the first assessment of ML-based tool usability performed in situ via participant observation.

Previous work on ML and visualization tool development includes tools such as Ocelot~\cite{arendt2015ocelot}, which was designed to help analysts make better decisions about poorly defined network intrusion events, 
Situ~\cite{goodall2018situ}, used to identify anomalous behavior in network traffic, 
and the work of Best et al.~\cite{best2010real}, which seeks to give analysts situational understanding of the network utilizing complementary visualization techniques. 
Bridges et al.~\cite{bridges2018forming} introduced the Interactive Data Exploration \& Analysis System (IDEAS), a research prototype allowing analysts to query data in their SOC log store and select ML models to be run ``under the hood'', then receive outputs in an interactive visualization.
Sopan et al.~\cite{sopan2018building} generated a machine learning model to aid SOC analysts in isolating meaningful alerts by conducting two hour interviews with the five most experienced analysts in the SOC to better understand their workflow. 
They then created a prediction explanation visualization to aid analysts and stakeholders in understanding how the model was making decisions.
%They did not do a usability study, but they did receive feedback from the analysts indicating that they found the score useful. 
%They also found that their initial user interface, which assumed a base level of knowledge about machine learning, had to be modified for the analysts who were not as familiar with relevant terminology. 
%Their work highlights the importance of making machine learning comprehensible to analysts in order to reap its benefits.  

Work in the second category considers methods for evaluating the effectiveness of security tools. 
Akinrolabu et al.~\cite{akinrolabu2018challenge} interviewed expert SOC analysts to better understand obstacles to detecting sophisticated attacks and Cashman~\cite{cashman2019user} conducted a user study of a novel approach to developing machine learning models that involved users in the selection process.
They both suggest that involving the user in the creation of the machine learning model can provide significant benefits.
Jaferian et al.~\cite{jaferian2014heuristics} proposed a new set of usability heuristics based on activity theory that would complement rather than replace traditional methods such as Nielsen's heuristics. 

Work in the third category focuses on understanding SOC operators. 
Gutzwiller et al.~\cite{gutzwiller2016task} performed a cognitive task analysis to understand the goals and abstracted elements of awareness cyber analysts use in their jobs. 
They found that data fusion in visualizations is most useful when it is combined with a strong knowledge of the network itself on the part of the analyst. 
%They called this form of awareness cyber-cognitive situation awareness (CCSA) and suggested it is a critical intermediary that allows analysts to use the tools at their disposal effectively. 
These results match findings by Ben-Asher et al.~\cite{ben2015effects} that suggest situated knowledge about a network is necessary to make accurate decisions. 

Botta et al.~\cite{botta2007towards} interviewed a dozen SOC analysts in five companies and found that inferential analysis, pattern recognition and what they call ``bricolage'', or construction with whatever is at hand, are key skills for IT security professionals. 
Sundaramurthy et al.~\cite{sundaramurthy2016turning} conducted a 3.5 year long anthropological study of four academic and corporate SOCs and concluded that the only way to get new tools incorporated into existing workflows is to meet the spoken and unspoken requirements of analysts and their managers. 
In a previous study~\cite{sundaramurthy2015human}, they also developed a model for understanding SOC analyst burnout. 
Goodall et al.~\cite{goodall2004work}, Bridges et al.~\cite{bridges2018information}, and Kokulu et al.~\cite{kokulu2019matched} conducted interviews with security analysts to better understand SOC workflows and the problems plaguing SOC operations. 
Common problems include disagreements between managers and analysts and low visibility into network infrastructure and endpoints. 
%However, in contrast to prior work, Kokulu et al.~\cite{kokulu2019matched} found that false positives in malicious activity detection do not majorly impact SOC operations.

Work in the fourth category is on ML for cybersecurity. 
As discussed by the position paper of Sommer and Paxon~\cite{sommer2010outside}, many pitfalls exist when applying  machine learning to cybersecurity\textemdash most notably, the ``semantic-gap'', referring to the common difficulty of analysts understanding the output of ML algorithms. 
The challenge is presenting results in a context that is understandable to, and actionable by, the analysts. 
More generally, the role of humans interacting with machine learning (ML) systems and the related usability challenges are areas of open research~\cite{gillies2016human}. 
There is also a plethora of work on the interpretation of ML algorithms, but we do not have space to include it.
For a summary, see the work of Gilpin et al.~\cite{gilpin2018explaining}.
%Related works fall into three categories---methods to evaluate the effectiveness of security tools in the context of a SOC~\cite{akinrolabu2018challenge,cashman2019user,jaferian2014heuristics}, studies on SOC operations~\cite{botta2007towards,sundaramurthy2016turning,sundaramurthy2015human,goodall2004work,bridges2018information,kokulu2019matched}, and ML for cybersecurity~\cite{sommer2010outside,gillies2016human,gilpin2018explaining}. 
%While prior work relied heavily on interviews or surveys for data collection, our work represents the first assessment of ML-based tool usability performed in situ via participant observation.

\section{Methodology}
In this section, we discuss our study design, data analysis, and demographics.

\subsection{Study Design}
This study was not comparative, but rather exploratory in nature.
Our goal in this work was to identify usability concerns in ML-based tools; not to compare the efficacy of the two tools being tested.
In order to achieve this goal, we observed participants during tool usage, administered a follow-up survey, and held a focus group to better understand users' experience.
We used the \emph{think-aloud} methodology~\cite{van1994think} during observation, in which participants verbalized their intentions, so that researchers would be able to understand the reasons behind participant actions.
By conducting the focus group after direct observation of each analyst, we utilized it as a way to supplement and refine our observations rather than as a sole source of data~\cite{nielsen1997use,nielsen1994usability}.

The participant observation consisted of two campaigns, one for each tool, in which we performed a sequence of  malicious actions against the network and analysts utilized the user interface provided by the tool to attempt to gain insight into the attack. 
Each campaign lasted one hour and fifteen minutes. 
Prior to the campaign, analysts were given an introduction to each tool and time to familiarize themselves with the interface. 
During this familiarization period, analysts could ask any questions they had regarding usage of the tool. 
Answers were directed to the entire group. 

During each campaign, the same researcher was assigned to each analyst to record information about and observe the analyst's use of the tool. 
An additional researcher was responsible for monitoring network status and providing notices every fifteen minutes.
Think-aloud was practiced during the familiarization period to ensure analysts understood it. 

Analysts also recorded insights from each tool they thought were significant as they used the tool and rated the significance of each insight. 
Following each test, analysts were surveyed to better understand their experience with the tool and the observers were able to ask for any necessary clarification. 
The survey included the System Usability Scale (SUS) along with additional questions designed by the researchers. 
The day after testing, we held a focus group to supplement and refine our observations.

\subsection{Attack Campaigns}
\label{sub:red}
We created an attack campaign template that contained actions that one or both of the tools under test should catch.  
During each testing period, we ran through the actions specified in the attack campaign template while slightly permuting the IPs and payloads used so that the analysts' experience from one tool test would not impact their results in the next.
Generally, the attack campaigns consisted of the following actions.

First, the adversary gains initial access by dropping a customized version of Cobalt Strike's  Beacon\footnote{\url{https://www.cobaltstrike.com//help-beacon/}}, a program mimicking APT's in allowing external access, on the initial target. 
This was meant to simulate a successful phishing attack, wherein an unsuspecting user of the target system is tricked into downloading and running a malicious email attachment.
From the infected target, the adversary port scanned other hosts on the network of the first compromised system.
The adversary then instructed the infected system to download additional malware over HTTP and then transfer the malware to another host on the network over Samba.
The adversary then ascertained administrator credentials by using Beacon's Hashdump functionality. 
With the newly found administrator privileges, the adversary used \texttt{PSEXEC} to laterally move from the infected foothold to another target on its internal network.
The adversary then exfiltrated some data from the file system of the newly infected host back to the command and control server (C2) and disconnected from the infected target.

\subsection{Data Analysis}
\label{sub:analysis}
Our data analysis was broken down into quantitative and qualitative components. 
The System Usability Scale (SUS) and attacks detected by each analyst were quantitative metrics, while the post-test survey and focus group were qualitative. 
For the qualitative analysis, we used a modified version of the open coding approach~\cite{strauss1998basics} called pair coding~\cite{sarker2000building,salinger2008coding}, in which researchers create and assign codes collectively.

For the follow-up survey, we also conducted a sentiment analysis. 
Each coder counted $p$, the number of positive, and $n$, the number of negative comments, for each question. 
We report and define $S_r: = (p-n) / (p+n)$, a sentiment ratio. 
Note that $S_r \in [-1,1]$ with $S_r = \pm 1$ if all comments were positive/negative, respectively, and $S_r = 0$ if the quantity of positive and negative comments were equal.
We added the $p$ and $n$ values of both researchers together and then calculated a composite sentiment ratio. 

%\subsection{}

%In this preliminary work we only considered two tools.
%We plan to examine a broader range of tools in future work.
%While our sample size was small, we believe our results are transferable to other analysts with comparable backgrounds~\cite{krefting1991rigor}.
%However, analysts' lack of familiarity with the network under test may have impacted their ability to make accurate inferences from the tools~\cite{gutzwiller2016task,ben2015effects}.

\subsection{Recruitment \& Ethics}
This IRB-approved study was conducted as part of a tool evaluation exercise organized by our Navy sponsor.
In order to participate, analysts were required to be actively employed in one of the sponsor's SOCs. 
The sponsor provided six analysts for the event, with both experienced and novice analysts included in the sample.
Prior to testing, we went over an information sheet detailing the nature of the research and the participants' rights.

\subsection{Demographics}
%Table~\ref{tab:analystdata} contains demographic data for each analyst. 
Half of the analysts' highest level of education was high school, while two had completed a Bachelor's and one an Associate's degree. 
For context, most IT security professionals have either a Bachelor's or an Associate's degree.~\footnote{\url{https://itcareercentral.com/security-roles-salary-expectations-explained/}}
Half of the analysts had one year or less of experience on the job, while the others had three, eight, and five years of experience. 
Ages ranged from twenty-six to thirty-seven.  
Table~\ref{tab:analysttools} shows the tools each analyst reported using on their job regularly. 

%\begin{table}[h]
%\centering 
%\begin{tabular}{@{}clccc@{}}
%\toprule
%\textbf{Analyst} & \textbf{Education} & %\textbf{Age} & \textbf{Experience}\\
%\midrule
% 1 & High School & 27 & 2 months \\ 
% 2 & Bachelor's  & 32 & 3 years\\
% 3 & Associate   & 26 & 1 year\\
% 4 & High School & 28 & 1 year\\
% 5 & Bachelor's  & 37 & 8 years\\
% 6 & High School & 26 & 5 years\\ \bottomrule
%\end{tabular}
%\caption{Demographic Data}
%\label{tab:analystdata}
%\end{table}

\section{Analysis \& Results}
\label{sec:results}
In this section, we discuss our key findings and make recommendations for UI designers based upon the usability issues we identified.  
While our study is preliminary in nature, our findings demonstrate that ML-based security tool vendors must put a renewed focus on working with analysts, both experienced and inexperienced, to ensure that their systems are usable and useful in real-world security operations settings.

\subsection{Tool Usability}
To evaluate the overall usability of each tool, we used the System Usability Scale (SUS). 
For the SUS, ten statements are ranked from 1 to 5, where 1 is strongly disagree and 5 is strongly agree.
Half of the statements express a positive experience with the tool and half a negative experience with the tool.
The responses are then converted to a composite score on a scale from 0-100, where a score above 68 is considered average, an 81 would be an `A’ and a 50 would be an `F’.

\begin{figure*}[!ht]
\centering
\begin{subfigure}{.25\textwidth}
  \centering
  \includegraphics[width=45mm]{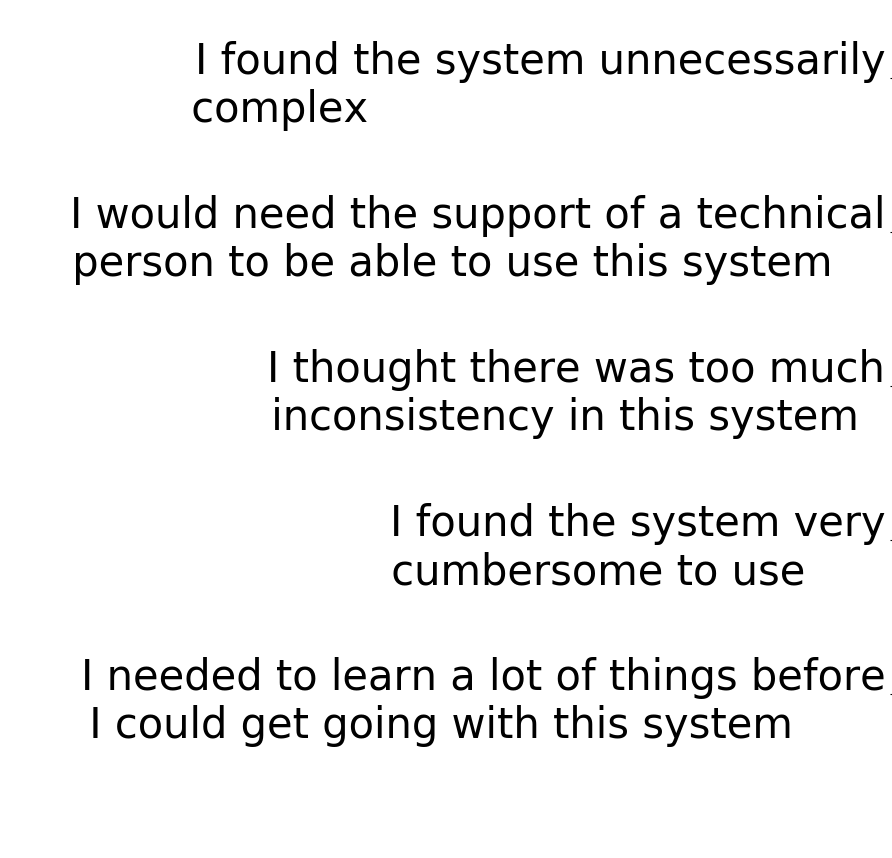}
  \label{fig:sub0}
\end{subfigure}
\begin{subfigure}{.33\textwidth}
  \centering
  \includegraphics[width=53mm]{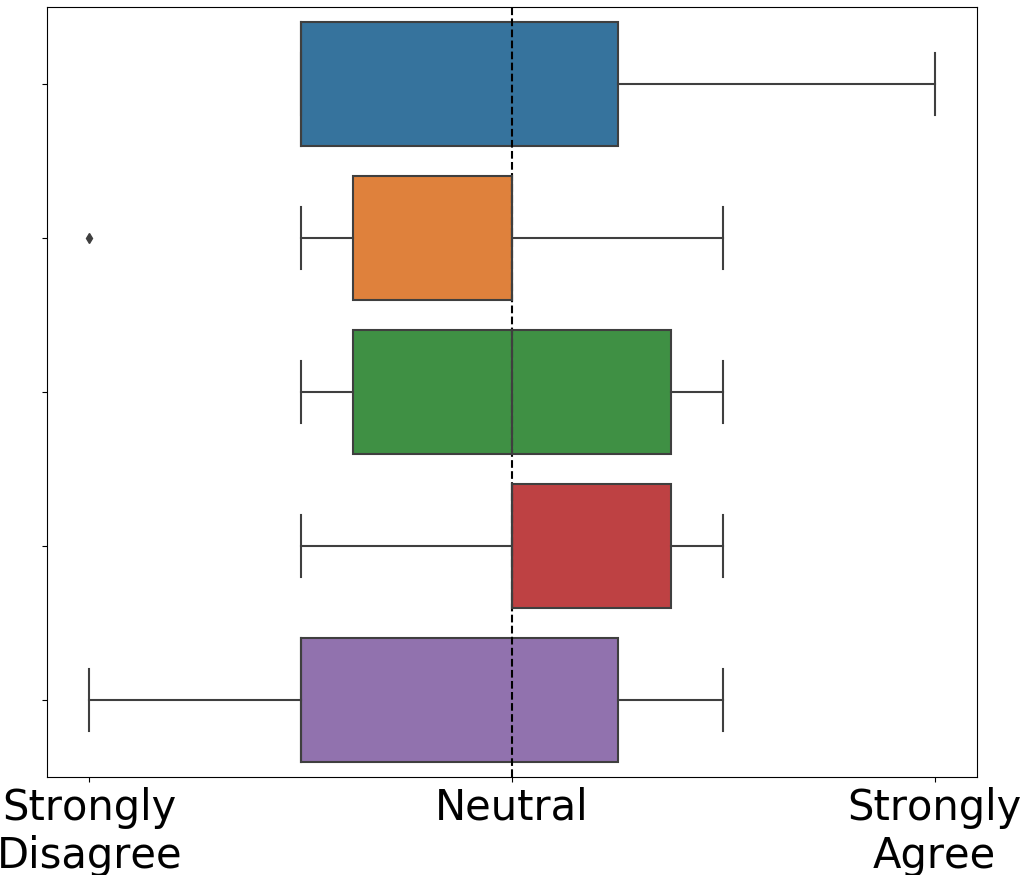}
  \caption{NSDT}
  \label{fig:sub3}
\end{subfigure}
\begin{subfigure}{.33\textwidth}
  \centering
  \includegraphics[width=53mm]{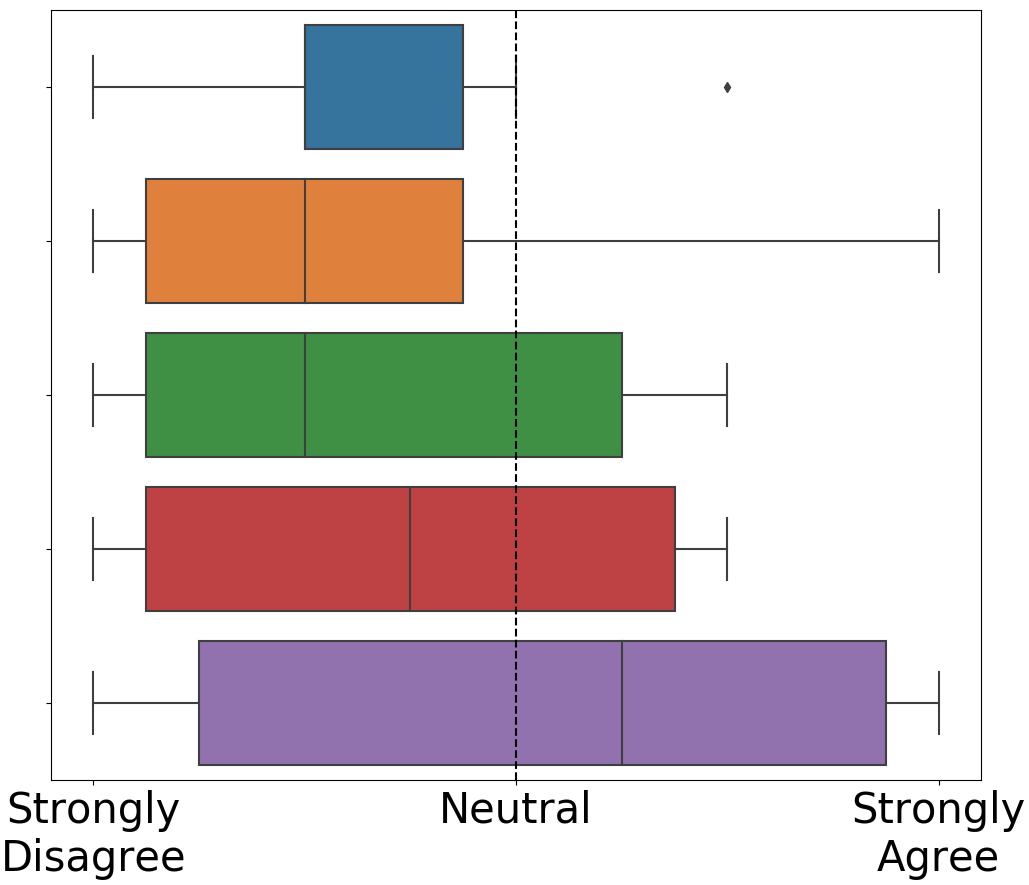}
  \caption{Situ}
  \label{fig:sub4}
\end{subfigure}
\caption{SUS Statements Expressing a Negative Experience}
\label{fig:negsent}
\end{figure*}

\begin{figure*}[!ht]
\centering
\begin{subfigure}{.25\textwidth}
  \centering
  \includegraphics[width=45mm]{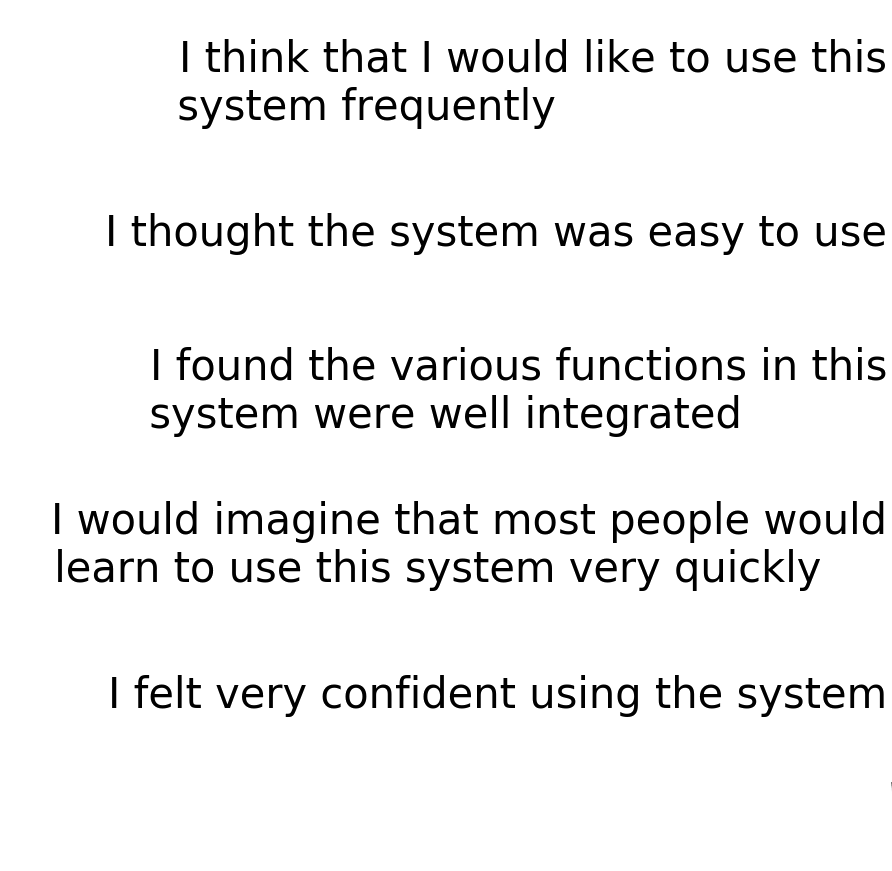}
  \label{fig:sub}
\end{subfigure}
\begin{subfigure}{.33\textwidth}
  \centering
  \includegraphics[width=53mm]{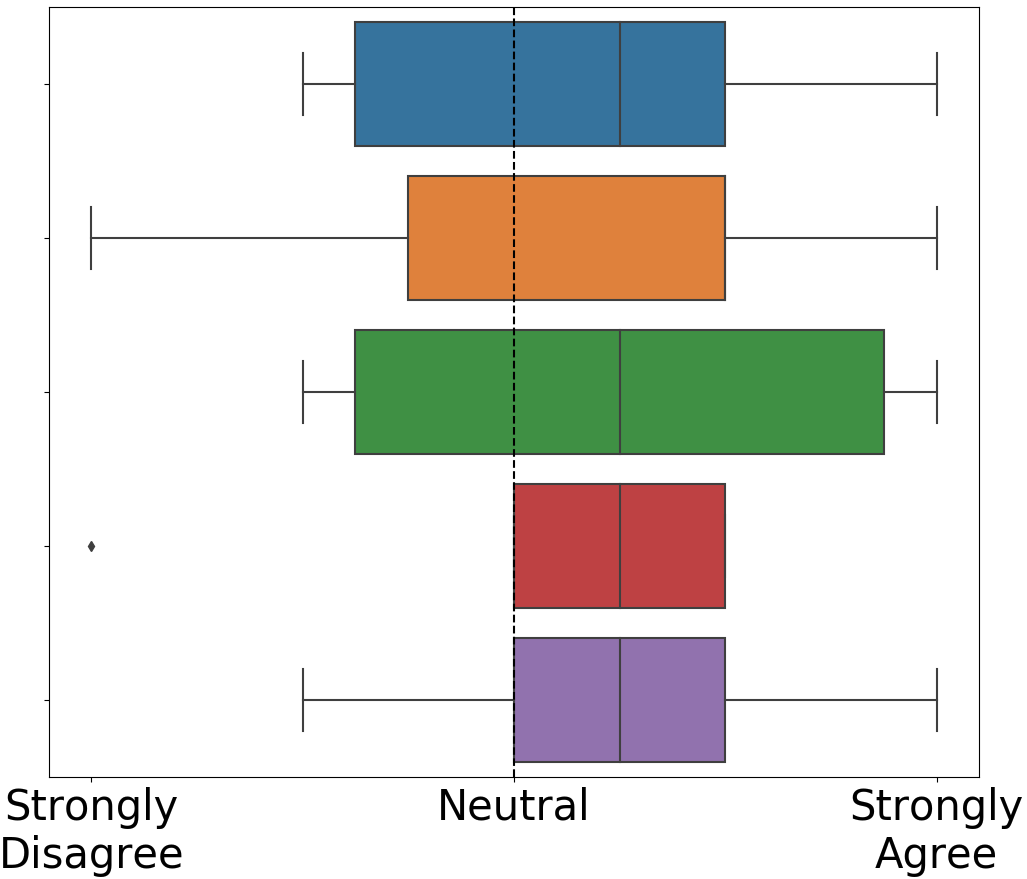}
  \caption{NSDT}
  \label{fig:sub1}
\end{subfigure}
\begin{subfigure}{.33\textwidth}
  \centering
  \includegraphics[width=53mm]{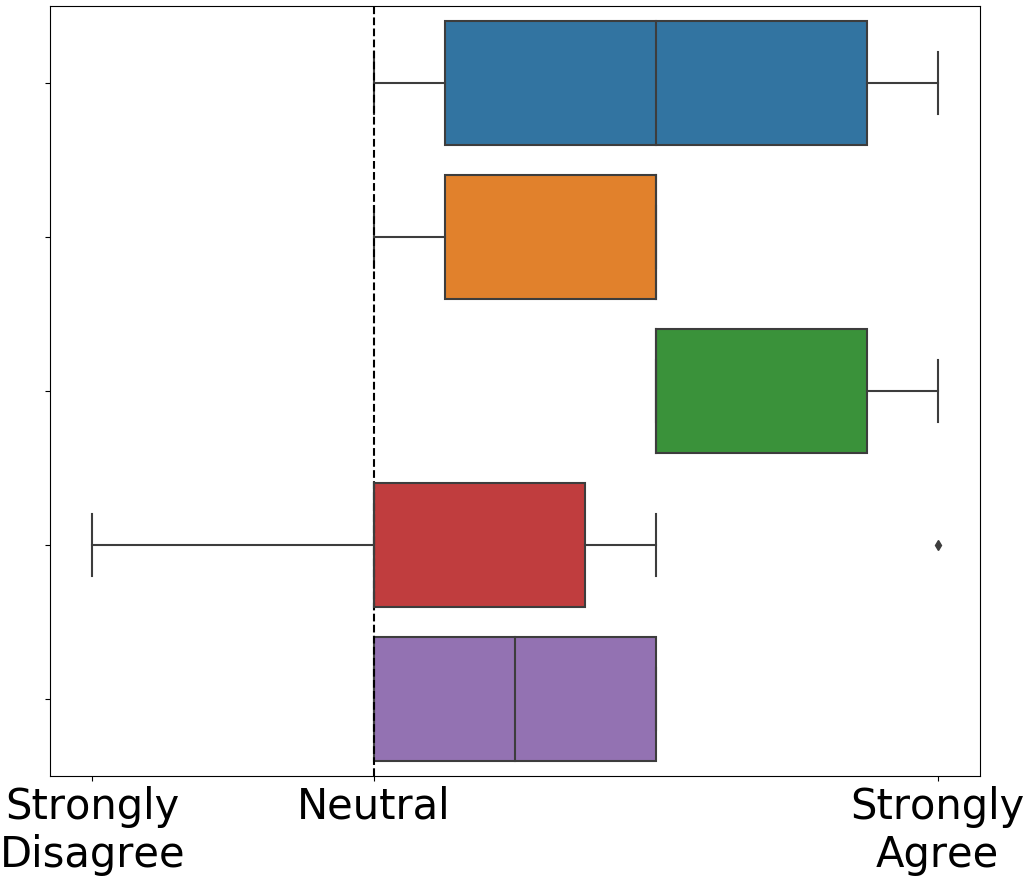}
  \caption{Situ}
  \label{fig:sub2}
\end{subfigure}
\caption{SUS Statements Expressing a Positive Experience}
\label{fig:possent}
\end{figure*}

The SUS results for the statements expressing a negative experience are shown in Figure~\ref{fig:negsent} and the results for the statements expressing a positive experience in Figure~\ref{fig:possent}.
The mean score for Situ was 65.42, which is average, while NSDT was closer to the failure line with a 56.67.
Given that NSDT is a commercially available tool, this result is disappointing. 
Analysts indicated that NSDT is cumbersome and that it contained inconsistencies, issues we will see again in the next section.
For Situ, the main issue identified by the SUS was that analysts felt they needed to learn a lot before they could use the system effectively. 
We suspect analysts responded this way to Situ for two reasons.
First, Situ required analysts to synthesize multiple views of the same data built on different statistics (anomaly score, PCR, geographic information).
Second, Situ identified anomalous, rather than malicious, activity, requiring analysts to decide when anomalous behavior was worth investigating.
The fact that most analysts lacked a clear mental model for how to use the anomaly scores presented by Situ, which we will discuss in Section~\ref{sub:mental}, supports this explanation. 

To verify that these results were approaching saturation (i.e. they would not change substantially even if we added more analysts), we also computed the hold-one-out average scores with only five of the six analysts for all six combinations.
This yielded six average scores: 62.0, 62.5, 63.5, 65.5, 67.5, and 71.5 for Situ and 50.0, 55.0, 54.5, 58.0, 59.0, and 63.5 for NSDT.
The similarity in these average scores verifies that our SUS results are near saturation.

\begin{table}[t]
	\setuptable
	\begin{tabular}{c|ccccccc}
		\multicolumn{1}{l}{Analyst} \headline[4.5cm]
	    & \headrow{Network Analysis Framework} 
		& \headrow{Automated Malware Analysis} 
		& \headrow{Network Packet Analyzer} 
		& \headrow{Putty, Bash or Powershell}
		& \headrow{Full Stack Analytics}
		& \headrow{SIEM} 
		& \headrow{IDS} 
		\\ 
		\hline
		
		1		&\none			&\none		 &\full	              &\none          &\none	     &\full		        &\none \\
		
		2		&\full			&\none		 &\full	              &\full          &\full	     &\none		        &\full \\
		
		3		&\none			&\full		 &\full	              &\full          &\none	     &\full		        &\none \\
		
		4		&\full			&\full		 &\full	              &\none          &\none	     &\none		        &\full \\
		
		5		&\full			&\full		 &\full	              &\full          &\none	     &\full		        &\full \\
		
		6		&\full			&\none		 &\none	              &\none          &\none	     &\full		        &\full \\
    \hline
	\end{tabular}

	\caption{Tools Analysts Reported Using Regularly}
	\label{tab:analysttools}
\end{table}
  
\subsection{User Interface Issues}
Table~\ref{tab:heuristics} summarizes which of Nielsen's heuristics~\footnote{\url{https://www.nngroup.com/articles/ten-usability-heuristics/}} for user interface design each system violated.

With NSDT, analysts felt particularly frustrated by a lack of consistency in the user interface.
Multiple pages contained overlapping content and looked similar, which caused analysts to continually feel lost because they were trying to remember which page contained which content. 
Some content was also only available for certain file types, exacerbating this feeling of confusion.
A2 said they "fought the GUI the entire hour" and A1 said they "had to click around a lot---inconsistency".

Analysts main frustration with Situ was that the filters applied to the search bar were only visible in the URL and were not easily modifiable, forcing analysts to start a new search from scratch if they wanted to alter search parameters. 
A4 said he/she ``hated filters not listed except in the url''.

One issue both tools had in common is that they failed to provide the analysts with as much information as they wanted about the scores produced by the tool. 
Discussing the score provided by NSDT, A4 noted, ``It seems accurate but I would want more info on why it thinks it's malicious provided in more of a clean way''.
While Situ did provide explanations in the website documentation, some analysts found them difficult to understand. 
ML-based tools need to provide clear and easily accessible explanations for how the ML algorithm scores events.
Pop-ups explaining each score should be provided with links to additional reading for those analysts who want to go more in depth.

\begin{table}[t]
	\setuptable
	
	%\rowcolors{2}{gray!10}{}
	\begin{tabular}{|l|cc|}
		\hline
		Heuristic & NSDT & Situ \\ \hline
		
		%%%%%%%%%%%%%%%%%%%%%%%%%%%
		% App
		%%%%%%%%%%%%%%%%%%%%%%%%%%%
		Visibility of System Status		& \good & \bad	\\
		System Matches Real World		& \bad  & \bad	\\ 
		User Control and Freedom		& \good & \bad	\\
		Consistency and Standards		& \bad  & \good	\\
		Error Prevention		        & \good & \good	\\
		Recognition Not Recall		    & \bad  & \good	\\
		Flexibility and Efficiency		& \good & \good	\\ 
		Aesthetic and Minimalist		& \good & \good	\\
	    Help Users with Errors		    & \good & \good	\\						
		Help and Documentation		    & \bad  & \bad  \\ \hline
	\end{tabular}

	%\rowcolors{3}{}{gray!10}
	\begin{tabular}{ll}
		\good & No observed violations of this heuristic\\
		\bad  & Observed violations of this heuristic	
	\end{tabular}

	\caption{Summary of whether or not each system observed Nielsen's heuristics for user interface design.}
	\label{tab:heuristics}
\end{table}

\subsection{How Mental Models Impact Distrust and Misuse of Tools}~\label{sub:mental}

With both NSDT and Situ, some analysts distrusted and/or misused the tool because they had an incorrect mental model of how scores were generated.
NSDT scored malicious files on a scale of 1 to 10, where 1 meant that the file was benign and 10 that it was malicious.
While analysts had little trouble identifying malicious files using this score even if they did not understand how it was generated, the machine learning engine also provided a confidence level along with the score.
This confidence level was always 100\%, a fact that A4 found suspicious, saying, "Why trust this score?".
An unclear mental model of how NSDT generated the confidence level resulted in A4 mistrusting the tool because the confidence level was always the same.
This result supports prior work~\cite{dovsilovic2018explainable}, which found that analysts who did not understand the ML algorithms distrusted the scores they provided

Unlike NSDT, Situ produced an anomaly score based on the flow of network traffic.
A more anomalous flow received a higher score. 
Analysts had varying mental models for how Situ worked and therefore approached anomaly scores very differently.
For example, A4 focused on any anomaly scores above a particular value they deemed significant, but discounted events as insignificant if the number of bytes transmitted was small.
A5 would investigate which model contributed most heavily to the score, but mainly focused on IP associations.
And A6 understood that they should use the anomaly scores to identify a sequence of malicious actions composing a campaign, but they did not understand how to decide which anomalous activity warranted further investigation.

%A2 used the scores to try to identify what normal network behavior looked like, but conflated anomalous with malicious. 
%(Although, overall A2 found the tool very useful, and exhibited a strong understanding of the anomaly scores and other data statistics available in Situ.) 
%A1 focused on internal-to-internal traffic rather than internal to external traffic and searched for individual malicious events rather than trying to identify the sequence of events composing the red team campaign.
%A3 spent a lot of time exploring the interface, but disregarded the anomalous scores produced by the tool.
%A4 focused on any anomaly scores above a particular value they deemed significant, but discounted events as insignificant if the number of bytes transmitted was small.
%A5 would investigate which model contributed most heavily to the score, but mainly focused on IP associations.
%And A6 understood that they should use the anomaly scores to identify a sequence of malicious actions composing a campaign, but they did not understand how to decide which anomalous activity warranted further investigation.

In summary, analysts misused Situ for several reasons: (1) They did not understand the difference between anomalous and malicious, (2) They did not understand how to map anomaly scores to attacker actions, (3) They did not know how to prioritize anomalous events. 
Even though we explained how anomaly scores were calculated during the familiarization period prior to testing and allowed analysts to ask for clarification, only A2 claimed to understand how anomaly scores were calculated during the focus group. 
These results suggest that AD tools such as Situ may require a more accurate mental model of how scores are produced in order for analysts to use them properly because they require analysts to make complex inferences from the score and to differentiate between anomalous and malicious. 
In contrast, NSDT flagged files as malicious or non-malicious on a scale of 1 to 10 and would not necessarily require any understanding of the ML model to use effectively, though a lack of understanding can lead to distrust. 

\subsection{Experience, Tool Performance and Tool-Analyst Match}

To assess performance, we let $fc$ and $tc$ denote the number of false and true conclusions made by an analysts, respectively, where $fcr: = fc/(fc + tc)$.
A false conclusion occurred when an analyst thought they found malicious activity with a tool, and the activity was actually benign.
Table~\ref{tab:expact} shows the number of attack actions identified by each analyst and their false conclusion rate, .
We found that the mean false conclusion rate for analysts was .57 (std=.13) with Situ and .28 (std=.25) with NSDT.
%It makes sense that the AD tool would have a higher false conclusion rate than the AV tool because the AD tool requires additional analysis to confirm that an anomaly is actually malicious, whereas the AV tool only flags files it deems malicious.

We did not find that an analyst's experience level directly correlated to an ability to use the tools.
With NSDT, an analyst with only 1 year of experience (A3) performed as well as an analyst with 8 years of experience (A5).
For Situ, an analyst with only 2 months of experience (A1) performed as well as another analyst with 5 years of experience (A6) and better than an analyst with 8 years of experience (A5).
We used a scatter matrix to check for correlations between performance and other demographic data collected, such as education, but found none.
This result is surprising. 
We expected analysts with more experience and education to outperform junior analysts.

We also found that most analysts performed better with one tool or the other.
A1 and A2 performed well with Situ, but poorly with NSDT. 
A3 and A5 performed well with NSDT, but poorly with Situ.
This result may suggest a tool-analyst match, where individual analysts are predisposed to certain tool types.

\begin{table}[h!]
\begin{center}
\begin{tabular}{ll|cc|cc}
 &  & \multicolumn{2}{|c|}{Situ} & \multicolumn{2}{c}{NSDT} \\
\hline
Analyst & Experience & $tc$ & $fcr$ & $tc$ & $fcr$ \\
\hline
A1 & 2 months & 3 & .5  & 1 & 0\\
A2 & 3 years &  4 & .43 & 1 & .67\\
A3 & 1 year &   2 & .5  & 4 & .2\\
A4 & 1 year &   1 & .8  & 2 & .5\\
A5 & 8 years &  2 & .71 & 4 & 0\\
A6 & 5 years &  3 & .5  & 5 & .33\\
\end{tabular}
\end{center}
\caption{Analyst Experience and Performance metrics depicted.
A false conclusion occurred when an analyst thought they found malicious activity with a tool, but the activity was actually benign.
Because NSDT flagged malicious files, an $fcr$ of 0 was possible for analysts who focused solely on flagged files and did not attempt to draw further conclusions about the nature of the attack.}
\label{tab:expact}
\end{table}

\begin{table*}[hbt!]
\centering
\begin{tabular}{@{}rcc@{}}
\toprule
\textbf{} & \textbf{NSDT} & \textbf{Situ}\\
\midrule
What was your overall impression of the tool? & .39 & .68 \\
Was this tool easy and intuitive to use? & -.08 & .09\\
How do you see this tool fitting into your workflow? & .44 & .53\\
If this tool was in your current work environment, would you use it? & .83 & 1\\
What was your impression of the alerts raised by the tool? & .56 & .53\\
\midrule
Average Sentiment Ratio & \textbf{0.43} & \textbf{0.53} \\%\textbf{2.14} & \textbf{2.83}\\
 \bottomrule
\end{tabular}
\caption{Sentiment ratio, $S_r: = (p-n) / (p+n)$ with $p, n$ the number of positive/negative statements, on post-test questionnaire  reported.   
Note that $S_r \in [-1,1]$ with $S_r = \pm 1$ iff all comments were positive/negative, respectively, and $S_r = 0$ iff $p=n$. 
In spite of the concerns regarding intuitiveness of the alerts raised by the tools, analysts expressed overwhelmingly positive sentiment that they would use both tools if they were integrated into their work environment.}
\label{tab:sentiment}
\end{table*}

\subsection{User Attitudes}

Overall, analysts were optimistic about the capabilities these tools could provide.
The analysts liked Situ because it allowed them to discover a wide range of attacker actions \textit{during} an attack, whereas they felt most tools only allow them to respond \textit{after} the attack has already taken place.
After using Situ, A2 shared that it was "better than waiting for a light to turn red to do your job”.
While analysts viewed NSDT as a more retroactive tool, because it flagged malicious files rather than identifying anomalies, they also felt it could help them automate their workflow and conduct additional analysis.

Table~\ref{tab:sentiment} summarizes the results of our sentiment analysis of the follow-up survey for each tool, described in Section~\ref{sub:analysis}.
Analysts expressed a more positive overall impression of Situ than NSDT.
One possible explanation for this fact is that several analysts were very frustrated with NSDT's user interface for reasons noted in the previous section.
As a group, analysts did not find either tool particularly intuitive, expressing neutral sentiment for this question.
Analysts also showed some reservations about the alerts raised by the tools and how each tool would fit into their workflow.

In spite of these concerns, analysts expressed overwhelmingly positive sentiment that they would use both tools if they were integrated into their work environment. 
These results suggest that analysts are excited about the possibilities that ML tools provide and willing to use them in practice.
However, ML-based security tool vendors still have plenty of work to do to enhance the usability of their products, including addressing UI issues, helping analysts interpret alerts, and establishing a more intuitive workflow.

\section{Discussion \& Future Work}\label{discussion}
%Given the projected increase in the usage of ML-based tools in SOCs, it is vital for the research community to both enumerate and address usability concerns.
This work identified several serious usability issues in the two ML-based tools studied, including failure to follow established usability heuristics for user interface design and a lack of transparency into how scores are produced that caused distrust and/or misuse among analysts.
In light of these problems, we make the following recommendations:

\begin{enumerate}
    \item Vendors should conduct usability tests with actual SOC analysts, both experienced and inexperienced, throughout the software development life cycle.
    While heuristic evaluations are valuable, they require expertise to apply properly~\cite{thovtrup1991assessing} and are not as effective at identifying major issues pertinent to real users~\cite{paz2015heuristic}.
    This suggestion is also supported by the work of Bano et al.~\cite{bano2013user}, which concluded that software systems benefit from the inclusion of users in early stages of product development.
    \item ML-based tools should provide analysts with more guidance on how to understand and utilize their output.
    The benefit of ML is lost if analysts cannot understand the meaning of the scores produced.
    Prior research recommends including analysts when developing machine learning models to ensure interpretability~\cite{akinrolabu2018challenge,cashman2019user}.
    At a minimum, the vendor should conduct usability tests to validate that analysts are able to comprehend and use the scores produced by ML tools as intended by the vendor.
\end{enumerate}

The lack of sufficient explanation of ML concepts in either of the user interfaces we examined resonates with prior work.
Sopan et al.~\cite{sopan2018building} found that their initial user interface, which assumed a base level of knowledge about machine learning, had to be modified for the analysts who were not as familiar with relevant terminology.
Usable ML tools must bridge the ``semantic gap''~\cite{sommer2010outside} to help analysts who are not machine learning experts identify actionable insights. 

In addition, our results showed not only that incorrect mental models can cause distrust and misuse of tools, but also suggest that certain categories of ML tools require analysts to have more accurate mental models.
Specifically, we found that Situ, an AD tool, required a more accurate mental model to use because analysts had to make inferences based upon anomaly scores, whereas NSDT, an AV tool, flagged files as malicious or non-malicious and was therefore simple to interpret without any understanding of the underlying models.
While prior research has explored how mental models impact the usability of encryption~\cite{wu2018tree}, the Tor browser~\cite{winter2018tor}, and password managers~\cite{pearman2019people}, no research has focused specifically on how mental models impact SOC analysts' usage of ML-based tools.

Our research also uncovered the possibility of a tool-analyst match.
All analysts performed better with one tool or the other, yet we found no correlation between the demographic information we collected and performance.
These results suggest that other factors such as prior background knowledge or personality play a significant role in ML-based tool usage. 
While exploring personal attributes that impact tool usage was not the focus of our study, we believe this is an area that would be fruitful for researchers to explore further.

We plan to continue this work in several ways. 
First, we want to analyze a broader set of ML-based tools in order to identify usability paradigms and common issues within each paradigm. 
Second, we want to categorize analysts' mental models of different tool types and understand how those mental models impact their ability to use the tools.
The analysts in this study were excited about integrating ML tools into their SOCs and our research aims to help ensure that those tools are both usable and useful in real-world contexts.

%\input{60-limitations}
%\input{future.tex}
%\input{90-conclusion.tex}
%\input{97-acks.tex}

%-------------------------------------------------------------------------------
%\section*{Availability}
%-------------------------------------------------------------------------------

%USENIX program committees give extra points to submissions that are
%backed by artifacts that are publicly available. If you made your code
%or data available, it's worth mentioning this fact in a dedicated
%section.

%-------------------------------------------------------------------------------

\section*{Acknowledgment}
The research is based upon work supported by the Department of Defense (DOD), Naval Information Warfare Systems Command (NAVWAR), via the Department of Energy (DOE) under contract  DE-AC05-00OR22725. The views and conclusions contained herein are those of the authors and should not be interpreted as representing the official policies or endorsements, either expressed or implied, of the DOD, NAVWAR, or the U.S. Government. The U.S. Government is authorized to reproduce and distribute reprints for Governmental purposes notwithstanding any copyright annotation thereon.

\bibliographystyle{plain}
\bibliography{references.bib}

%%%%%%%%%%%%%%%%%%%%%%%%%%%%%%%%%%%%%%%%%%%%%%%%%%%%%%%%%%%%%%%%%%%%%%%%%%%%%%%%
\end{document}